\def\la{\hbox{{\lower -2.5pt\hbox{$<$}}\hskip -8pt\raise
-2.5pt\hbox{$\sim$}}}
\def\ga{\hbox{{\lower -2.5pt\hbox{$>$}}\hskip -8pt\raise
-2.5pt\hbox{$\sim$}}}
\def\ltsima{$\; \buildrel < \over \sim \;$}
\def\simlt{\lower.5ex\hbox{\ltsima}}
\def\gtsima{$\; \buildrel > \over \sim \;$}
\def\simgt{\lower.5ex\hbox{\gtsima}}

\documentclass[12pt,manuscript]{aastex}

\begin{document}


\title{GRB941017: A Case Study of Neutrino Production in Gamma Ray Bursts}


\author{Jaime Alvarez-Mu\~niz} 
\affil{Departamento de F\'\i sica de Part\'\i culas, Facultade de F\'\i sica, \\
15706 Santiago de Compostela, A Coru\~na, Spain}

\and

\author{Francis Halzen}
\affil{University of Wisconsin, Department of Physics,\\ 
1150 University Avenue, Madison, WI 53706, USA}

\and

\author{Dan Hooper}
\affil{Denys Wilkinson Laboratory, Astrophysics Department, \\ 
OX1 3RH Oxford, England UK}


\begin{abstract}

GRB941017, a gamma-ray burst of exceptional fluence, has recently 
been shown to have a high-energy component which is not consistent 
with the standard fireball phenomenology. If this component is the 
result of photomeson interactions in the burst fireball, it provides 
new and compelling support for substantial high-energy neutrino fluxes 
from this and similar sources. In this letter, we consider what impact 
this new information has on the neutrino spectra of gamma-ray bursts 
and discuss how this new evidence impacts the prospects for detection 
of such events in next generation neutrino telescopes.

\end{abstract}

\section{Introduction}

Gamma-Ray Bursts (GRBs) are the most powerful objects in the universe,
typically emitting luminosities between $10^{50}\,$ and $\,10^{54}\,$
erg/s over 0.1-100 seconds. Their energetics suggest that they may be
the sources of the highest energy cosmic rays
\citep{Waxman:1995,Vietri:1995}. 
It has been pointed out that, if high-energy cosmic rays are accelerated 
in GRBs, photomeson interactions between accelerated protons and 
target photons are 
inevitable, producing very high energy neutrinos and gamma-rays
(see, for instance, Waxman \& Bahcall 1997). 
Independent of any association of cosmic rays with GRBs, observable 
neutrino rates are predicted in models where similar energy goes into 
the acceleration of electrons and protons in the expanding fireball.

In October of 1994, the Burst And Transient Source Experiment (BATSE)
and the Energetic Gamma-Ray Experiment Telescope (EGRET) each observed
an exceptionally powerful example of such an object. GRB941017 has the
eleventh highest fluence observed in the nine years of BATSE
observations. More interesting, however, is the fact that this burst
displays not only the typical synchotron-inverse Compton spectrum at
$\sim$30-1000 keV, but also shows a power-law high-energy component extending
at least to 200 MeV in energy (Gonz\'alez et al. 2003). Other than this
burst, EGRET has seen four GRBs at energies of $\sim 100\,$ MeV. 
These were each consistent with an extension of the synchotron-inverse 
Compton spectrum \citep{Dingus:2001}. Additionally, Milagrito has 
also observed evidence for emission near $\sim 100 \,$ GeV in one burst. 
This observation lacked the ability to reveal significant spectral information,
 however \citep{Atkins:2000}. In a burst-by-burst analysis of the 
complete BATSE catalogue, Guetta et al. (2003) identified GRB941017 
as the most powerful neutrino emitter with, unfortunately, 
no neutrino telescope to observe it in 1994. Although other explanations 
may be possible, GRB941017 provides the best evidence to date for high-energy 
proton interactions with source photons in GRBs.

If protons are accelerated to energies sufficient to produce the 
gamma-ray spectrum observed in GRB941017, high-energy neutrinos are 
a necessary consequence \citep{Waxman:1997}. Although no high-energy 
neutrino telescope of sufficient volume was operational in October 1994, 
it is interesting to consider the prospects for detection of a burst with 
similar characteristics in future experiments.

In this letter, we calculate the neutrino spectrum predicted for 
such an event, and discuss the prospects for detection in future 
experiments such as the kilometer scale neutrino telescope 
IceCube \citep{Ahrens:2003}. Our conclusions are: 
i) An event like this is likely to be observable with 
$\sim 0.5 - 5$ events from a single burst and, 
ii) a handful of events of this type can produce the 
diffuse flux of order 10 events per year predicted by 
fireball phenomenology \citep{Guetta:2003}.


\section{Neutrinos From Photomeson Interactions In GRB Fireballs}

Accelerated protons in GRB fireballs produce parent pions via the processes
\begin{eqnarray}
p\gamma \rightarrow \Delta \rightarrow n \pi^{+}
\end{eqnarray}
and,
\begin{eqnarray}
p\gamma \rightarrow \Delta \rightarrow p \pi^{0}
\end{eqnarray}
which have very large cross sections of 
$\sigma_{\Delta}\sim 5\times 10^{-28}~\rm{cm}^2$.  
The charged $\pi$'s subsequently decay producing charged
leptons and neutrinos, while the neutral $\pi$'s decay into 
high-energy photons. To have sufficient center-of-mass energy 
for these processes to take place, protons must meet the 
threshold condition:
\begin{eqnarray}
\varepsilon^\prime_p \geq \frac{m_\Delta^2-m_p^2}{4 \varepsilon^\prime_\gamma}.
\end{eqnarray}
Primed and unprimed quantities refer to values measured in the 
comoving and observer's frames, respectively. 
In the observer's frame,
\begin{equation}
\label{Eq:epb}
\varepsilon_p \geq 1.4 \times 10^{16} \frac{\Gamma^2_{2.5}}
{\varepsilon_{\gamma,{\rm MeV}}}\rm{eV},
\end{equation}
resulting in a neutrino energy
\begin{equation}
\varepsilon_{\nu}= \frac{1}{4} \langle x_{p \rightarrow \pi} \
\rangle \varepsilon_p \geq 7 \times
10^{14}\frac{\Gamma^2_{2.5}}
{\varepsilon_{\gamma,{\rm MeV}}}\rm{eV},
\end{equation}
where $\Gamma_{2.5}=\Gamma/10^{2.5}$ is the bulk 
Lorentz factor and $\varepsilon_{\gamma, \rm{MeV}}=
\varepsilon_{\gamma}/~\rm{1 MeV}$ is the 
typical target photon energy. 
$\langle x_{p \rightarrow \pi} \rangle \simeq 0.2$ 
is the average fraction of energy transferred from the initial proton to the
produced pion. The factor of 1/4 comes from the assumption that the 4
final state leptons in the decay chain
$\pi^{+} \rightarrow \nu_{\mu} \mu^+ \rightarrow\nu_{\mu} e^+
\nu_e \bar{\nu_{\mu}}$ equally share the pion energy.

Typical GRBs display a broken power-law spectrum consistent with 
synchotron and inverse Compton emission. This spectrum can be 
parameterized by:
\begin{equation}
F_{\gamma}=\varepsilon_{\gamma}dn_{\gamma}/d\varepsilon_{\gamma}
\propto\left\{ \begin{array}{ll} 
\varepsilon_{\gamma}^{-\alpha} ;& 
\varepsilon_{\gamma}<\varepsilon^b_{\gamma} \\ 
\varepsilon_{\gamma}^{-\beta} ;& \varepsilon_{\gamma}>\varepsilon^b_{\gamma}
\end{array}
\right. \;.
\label{eq:Fnu}
\end{equation}
In the case of GRB941017, $\alpha$ and $\beta$ appear to change with time 
varying between 1.2 and 2.1 for $\beta$ and between $-0.2$ and $0.5$ 
for $\alpha$. In the first tens of seconds of the burst, the break 
energy is about 500 keV. It steadily drops, falling well below 
100 keV after about 100 seconds. 

For each proton energy, the resulting neutrino spectrum traces the GRB
photon spectrum, but with a much higher energy break:
\begin{equation}
\label{eq:enub}
\varepsilon^b_\nu = 
7 \times 10^{14}\frac{1}{(1+z)^{2}}\frac{\Gamma^2_{2.5}}
{\varepsilon^b_{\gamma,{\rm MeV}}}\rm{eV}.
\end{equation}
$z$ is the redshift of the GRB. For this particular burst, 
the redshift has been estimated to be $z \sim 0.6$, 
although its precise value is uncertain \citep{Guetta:2003}.

At very high energies, pions lose energy by synchrotron emission 
before decaying. This affects the spectrum of neutrinos from pion 
decay above the energy:
\begin{equation}
\label{eq:synclos}
\varepsilon^s_{\nu_{\mu}}=\frac{ 10^{17}}{1+z}\,
L_{\gamma,52}^{-1/2}\Gamma_{2.5}^4 t_{v,-2} \,~~ \rm{eV},
\end{equation}
where $L_{\gamma,52}$ is the gamma-ray luminosity of the burst in 
units of $10^{52}$ erg/s. For GRB941017, $L_{\gamma,52} \simeq 1$. 
$t_{v,-2}$ is the time scale of fluctuations in the GRB lightcurve 
in units of $10^{-2}~{\rm s}$
Neutrinos from muon decay have an energy cutoff that is 10 times
smaller due to the longer lifetime of the muon compared to that 
of the pion. Above this energy, the slope of the neutrino spectrum 
steepens by two.

The fraction of energy in accelerated protons converted to pions in the 
fireball is estimated from the ratio of the size of the shock, 
$\Delta R^\prime$, and the proton mean free path: 
\begin{eqnarray}
f_{\pi} \simeq \frac{\Delta R^\prime}{\lambda_{p \gamma}}\langle
x_{p\rightarrow \pi}\rangle.
\end{eqnarray}
Here, the proton mean free path is given by 
$\lambda_{p \gamma}=1/n_\gamma \sigma_{\Delta},$
where $n_\gamma$ is the number density of photons.
The photon number density is given by the
ratio of the photon energy density and the photon energy in the comoving
frame:
\begin{equation}
n_\gamma=\frac{U_\gamma^\prime}{\varepsilon_{\gamma}^\prime}\simeq
\bigg(\frac{L_\gamma t_v/\Gamma}{4\pi R^2 \Delta R^\prime}\bigg)
\bigg/
\bigg(\frac{\varepsilon_\gamma}{\Gamma}\bigg).
\end{equation}
Using these equations, and the relationship $R\simeq2 \Gamma^2 c t_v$, 
it is found
\begin{eqnarray}
n_\gamma\simeq\bigg(\frac{L_\gamma}{16\pi c^2 t_v
\Gamma^5 \Delta R^\prime  }\bigg)   \bigg/
\bigg(\frac{\varepsilon_\gamma}{\Gamma}\bigg)=
\frac{L_\gamma} {16\pi c^2 t_v\Gamma^4 \Delta R^\prime \varepsilon_{\gamma}},
\end{eqnarray}
which leads to
\begin{equation}
\label{eq:fpi1}
f_{\pi} \simeq
\frac{L_{\gamma}}{\varepsilon_{\gamma}}\frac{1}{\Gamma^4 t_v}
\frac{\sigma_\Delta \langle x_{p \rightarrow \pi} \rangle}{16 \pi c^2}
\sim 0.2 \times {L_{\gamma,52}\over
\Gamma_{2.5}^4 t_{v,-2}\varepsilon_{\gamma,\rm MeV}^b}.
\end{equation}
Thus far, this calculation describes protons at the break energy. 
In general,
\begin{equation}
\label{eq:fpi2}
f_\pi(\varepsilon_p)\sim 0.2~{L_{\gamma,52}\over
\Gamma_{2.5}^4 t_{v,-2}\varepsilon_{\gamma,\rm MeV}^b}
\times \left\{ \begin{array}{ll}
(\varepsilon_{p}/\varepsilon^b_{p})^{\alpha} &
\varepsilon_{p}>\varepsilon^b_{p} \\
(\varepsilon_{p}/\varepsilon^b_{p})^{\beta} & 
\varepsilon_{p}<\varepsilon^b_{p}
\end{array}
\right. \;,
\label{eq:fpi}
\end{equation}
where $\varepsilon^b_p$ is given by Eq.(\ref{Eq:epb}).

This equation indicates that $f_\pi$ and, therefore, 
the spectrum's normalization and the final event rates, 
vary significantly from burst-to-burst \citep{Halzen:1999,Alvarez:2000}. 
Such fluctuations are constrained, however, 
and similar conclusions are reached when fixing $f_\pi =0.2$, 
its typical value \citep{Guetta:2001}.

To estimate the bulk Lorentz factor, $\Gamma$, 
we relate it to the peak energy of the gamma-ray spectrum:
\begin{equation}
\label{eq:epeak}
\varepsilon_{\gamma}^b\approx 
\frac{ L^{-1/2}_{\gamma,52}}{\Gamma_{2.5}^2t_{v,-2}}~{\rm MeV}.
\end{equation}
For the case of GRB941017, we estimate 
$\Gamma \simeq$ 250 and this is again somewhat uncertain \citep{Guetta:2003}.

Finally, we obtain the neutrino spectrum:
\begin{eqnarray}
\label{Eq:nuflux}
\varepsilon^2_{\nu} \frac{dN_{\nu}}{d\varepsilon_{\nu}}  
\simeq \frac{1}{8} \frac{F_{\gamma}}
{\ln(10)} f_{\pi}
\end{eqnarray}
where, $F_{\gamma}$ and $f_{\pi}$ are determined 
by equations \ref{eq:Fnu} and \ref{eq:fpi2}, respectively. 
The neutrino spectrum  
found for GRB941017 in the time bins in (Gonz\'alez et al. 2003), 
is shown in figure 1.

\section{Event Rates in Neutrino Telescopes}

Currently, the AMANDA-II detector \citep{Andres:2001}, located at the 
South Pole, provides the strongest limits for high-energy neutrinos 
from GRBs \citep{Barouch:2001,Stamatikos:2003}. Considerably larger 
experiments such as the kilometer-scale neutrino observatory IceCube 
\citep{Ahrens:2003}, are presently under construction. For a review 
of high-energy neutrino astronomy, see Halzen \& Hooper 2002 or 
Learned \& Mannheim 2000.

High-energy neutrinos are observed as muons or showers created in 
interactions inside or near a Cherenkov detector embedded in an 
optically transparent medium such as ice or water. Muons, created 
in charged current interactions travel several kilometers before 
losing the majority of their energy, thus enhancing their prospects 
for observation. The probability for observing an energetic muon is
given by
\begin{equation}
P_{\nu_{\mu} \rightarrow \mu} = N_{\rm{A}} R_{\mu} \sigma_{\rm{CC}},
\end{equation}
where $N_{\rm{A}}$ is Avogadro's number, $R_{\mu}$ is the muon range 
\citep{Dutta:2001} and $\sigma_{\rm{CC}}$ is the neutrino-nucleon 
charged current cross section. For GRB941017, with a zenith angle of 
94.5$^{\circ}$ (for experiments at the South Pole), there is sufficient 
ice to allow for very long muon ranges.  In our calculations, we take 
into account the absorption of neutrinos in the Earth as well as the 
effect of oscillations. 

All three flavors of neutrinos may interact producing showers within 
the detector volume. The probability of making such an observation is 
similar to the relation for muons, although shower events do not benefit 
from long muon ranges, and typically they have to be produced inside the
detector to be observed. More details on how the rate is calculated 
can be found, for instance, in Appendix C of Guetta et al. 2003.

The rates for a GRB similar to GRB941017 are shown in table 1.
We have made two estimates of the rate. The first estimate uses 
Eq.~\ref{Eq:nuflux} for the neutrino flux, obtaining $F_\gamma$ from a 
broken power-law fit to the observed gamma-ray spectrum. It is important to 
remark that this fit cannot accommodate the late-time high-energy
feature of the gamma-ray spectrum.  
We predict 0.3 events per square kilometer, or roughly 1 event in 
IceCube with an effective area for muons and showers exceeding 1\,km\,$^2$.  
Bear in mind that for an observation over a narrow temporal and angular 
window in coincidence with the optical display, the background is 
entirely negligible \citep{Guetta:2003}. A single high-energy neutrino event 
is a conclusive observation. This rate is, of course, subject to the 
ambiguities associated with $z$ and $\Gamma$.  

To accommodate the late-time, high-energy feature seen in the gamma-ray 
spectrum, instead of using the method described above, we can, alternatively, 
estimate the neutrino flux and event rate by relating the observed 
high-energy gamma-ray component of GRB941017 to a corresponding high-energy 
neutrino component. This method assumes that the observed high-energy
gamma-ray component is the result of photomeson interactions. 
The observed high-energy gamma-ray component of GRB941017 can be parameterized 
(Gonz\'alez et al. 2003) by
\begin{equation}
\frac{dN_{\gamma}}{dE_{\gamma}} \simeq A_\gamma E_{\gamma}^{-\delta}
\end{equation}
where $\delta$ is measured to be 0.96 to 1.10 (Gonz\'alez et al. 2003) and
$A_\gamma$ is a normalization constant fixed to the observed spectrum. 
The corresponding 
neutrino flux can be calculated by energy conservation given that the physics 
of photomeson interactions fixes the ratio of energy going to photons and 
neutrinos \citep{Alvarez:2003}. The main uncertainties are the slope of 
the neutrino spectrum, and the energy to which the high-energy gamma-ray 
component extends. Without cascading of the gamma rays in the source both 
would be uniquely determined by the photon spectrum. This is not to be 
expected and we therefore calculate the neutrino flux varying the neutrino 
spectral slope and the high-energy cutoff of the gamma ray spectrum. Note 
that the latter is unobservable because high-energy gamma rays are absorbed 
on diffuse infrared background radiation.

The results are shown in figures 2 and 3. For a $E^{-1}$
neutrino spectrum and a gamma-ray cutoff near 100 TeV, we predict $\sim10$ events per square kilometer. Experiments capable of observing TeV 
gamma-rays from GRBs, such as MILAGRO \citep{Atkins:2001} and IceCube 
\citep{Halzen:2003} will be useful in determining the maximum energy 
to which the photomeson component of the gamma-ray spectrum extends.

Another way of summarizing our results is shown in figure 3 where the 
rate is plotted as a function of the neutrino spectral slope.  
The observed event rate increases with the spectral index because more 
neutrinos of lower energy are produced and, although their detection 
probability is smaller, their larger number compensates.
This trend continues until a neutrino spectral slope of $\sim 2.1$. 
Above this slope, a significant amount of the energy goes into
neutrinos below the experimental energy threshold of the detector and 
the rate decreases. The same argument explains why for a flatter
neutrino spectrum,
similar to the observed photon spectrum, the same event rate is reached
for a higher gamma-ray cutoff as seen in figure 2.

Although $\sim 10^3$ GRBs are observed per year, the total rate from 
all of these events is typically estimated to be $\sim 10$ events 
per square kilometer per year \citep{Guetta:2003}. A small number 
of exceptional 
bursts, such as the one we study here, can contribute substantially to 
the total neutrino flux.

\section{Conclusions}

We have pointed out that the high-energy feature observed in
GRB941017 provides further support for the expectation
of detectable fluxes of high-energy neutrinos in coincidence
with gamma ray bursts. We have estimated the neutrino spectrum
that would accompany such a burst and discussed the prospects for
such an event's detection in future neutrino observatories. We
find that such an event is expected to produce on the order of
1 event in a kilometer scale neutrino telescope and that this would be a conclusive observation since there is no competing background
during the time and in the direction of the burst.

\acknowledgements

J.~Alvarez-Mu\~niz is supported in part by Xunta de Galicia
(PGIDT02 PXIC 20611PN) and by MCyT (FPA 2001-3837 and FPA 2002-01161).
F.~Halzen work is supported in part by DOE grant No. DE-FG02-95ER40896,
NSF grant No.~OPP-0236449 and in part by the Wisconsin Alumni Research
Foundation.
D.~Hooper is supported by the Leverhulme Trust.

\newpage

\begin{figure}
\plotone{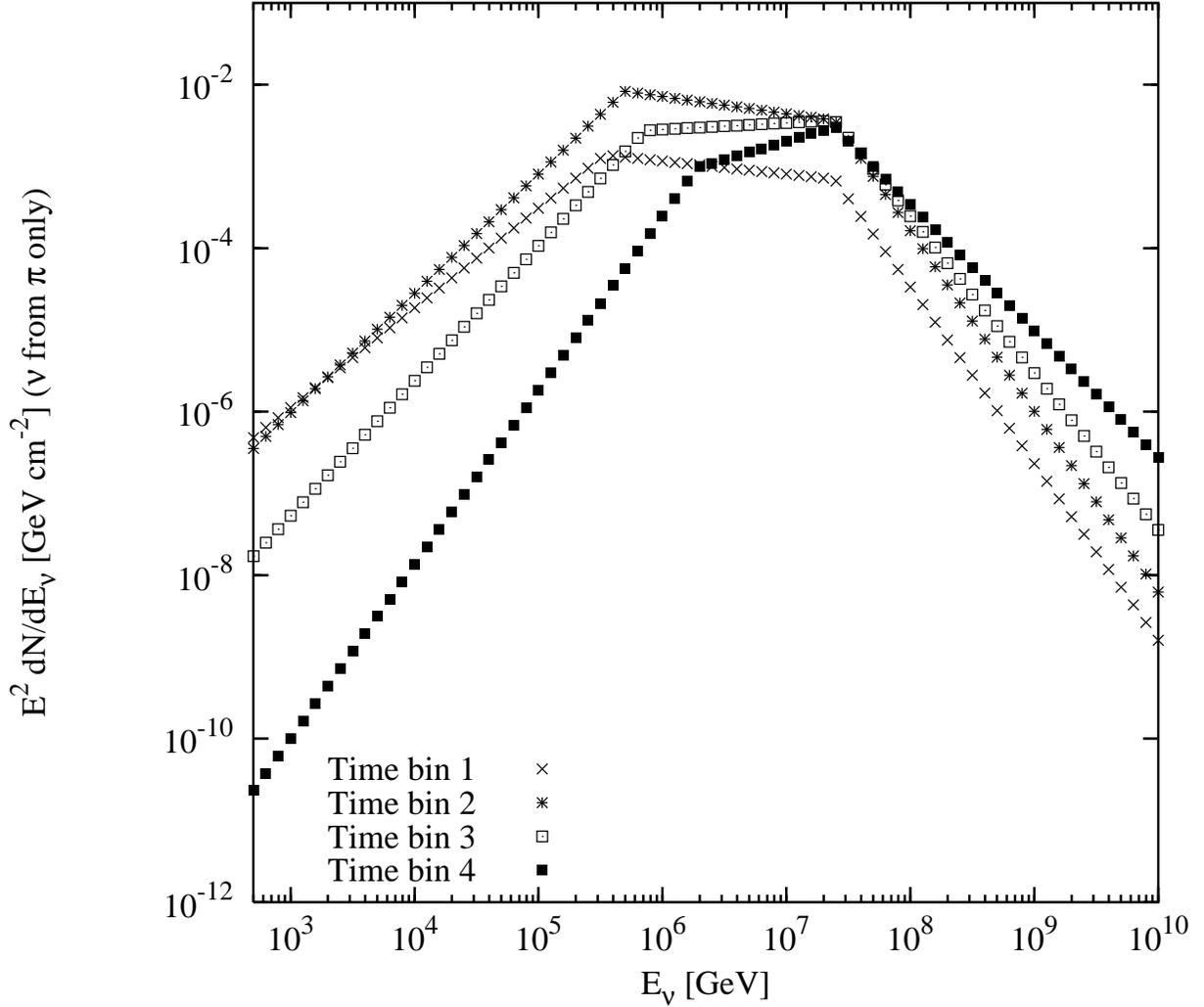}
\caption{The high-energy neutrino spectrum estimated for GRB941017. The
spectrum is shown for the four time bins in Gonz\'alez et al. 2003, 
the first of which begins 18 seconds prior to the BATSE trigger time. 
Each bin is of $\sim$33
second duration. In Gonz\'alez et al. 2003, a fifth time bin is described 
which we find to have a negligible impact on the neutrino spectrum.} 
\end{figure}

\begin{figure}
\plotone{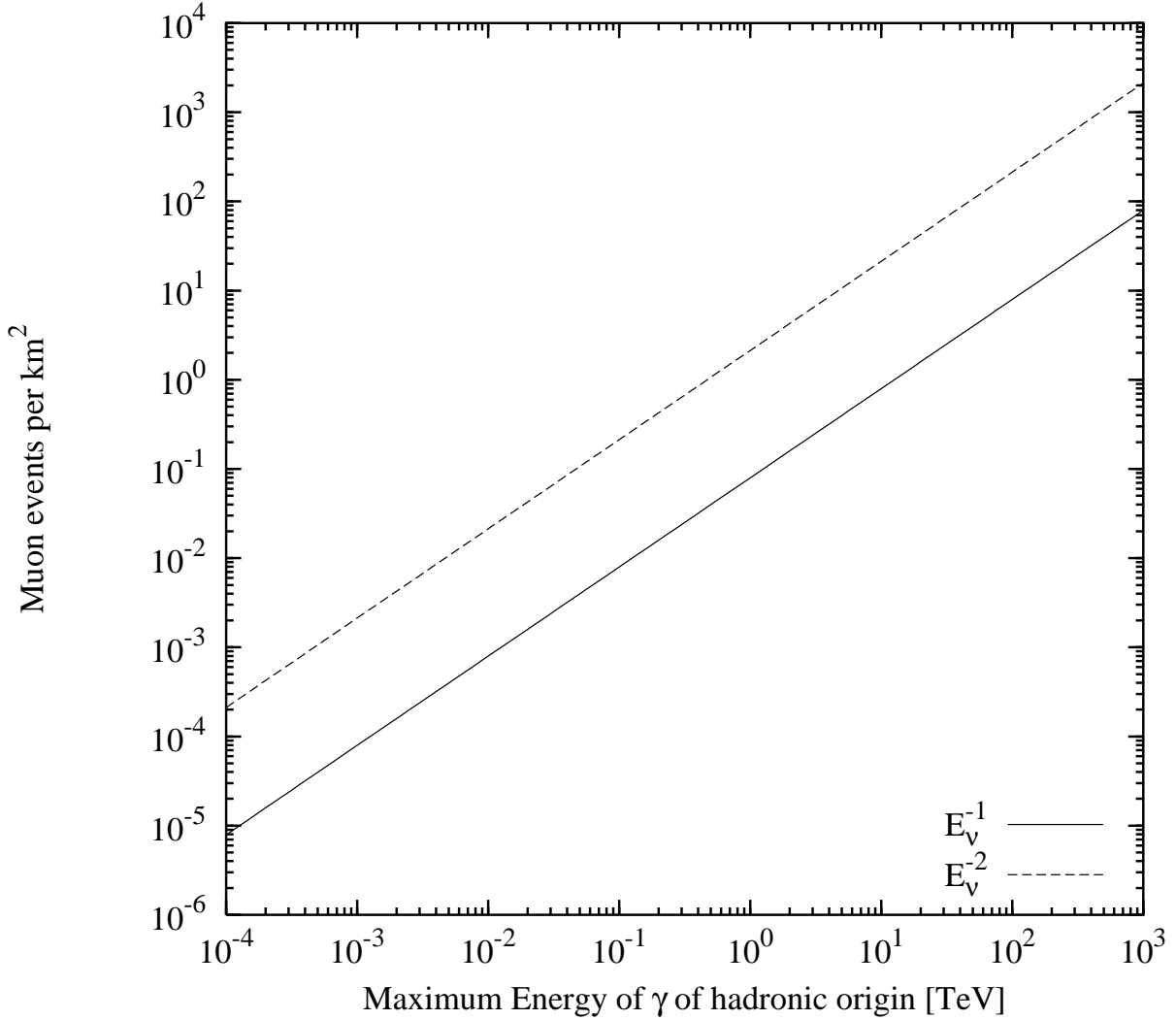}
\caption{The muon neutrino event rate per square kilometer 
in a high-energy neutrino 
telescope for GRB941017 as a function of the energy 
to which the high-energy gamma-ray component observed by EGRET extends.
The rate is calculated relating the observed high-energy gamma-ray 
component of GRB941017 to a corresponding high-energy neutrino component.  
Results are shown for two choices of accelerated neutrino spectra:
$E_\nu^{-1}$ (solid line), $E_\nu^{-2}$ (dashed line). Protons are assumed to 
be accelerated up to $E=10^{20}$ eV. The muon energy threshold is
$E_\mu^{\rm thr}$=100 GeV.}
\end{figure}

\begin{figure}
\plotone{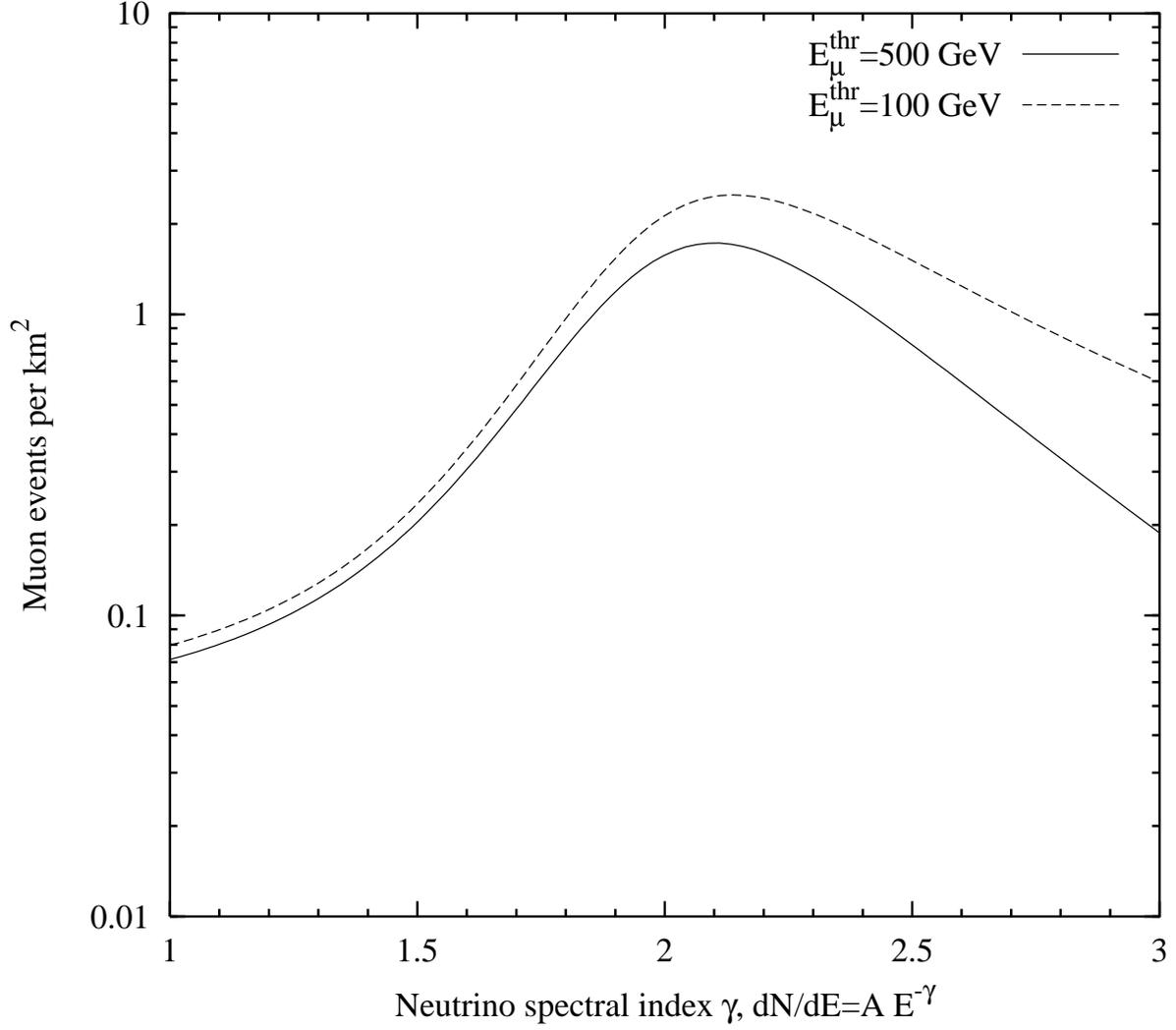}
\caption{The muon neutrino event rate per square kilometer in a high-energy 
neutrino telescope for GRB941017 calculated 
by relating the total energy of gamma-rays in the high-energy 
component to the total energy in high-energy neutrinos. Results 
assume the high-energy gamma-ray component observed by EGRET 
extends to 1 TeV. Protons are assumed to be accelerated up to 
$E=10^{20}$ eV. The effect of energy threshold is also shown: $E_\mu^{\rm thr}$=500 GeV (solid line); $E_\mu^{\rm thr}$=100 GeV (dashed line).}
\end{figure}

\vspace{0.0cm}
 \begin{table}
 \label{table}
 
 \hspace{1.5cm}
 \begin{tabular} {c c c c}
 \hline 
Calculation Method & Muons & Showers & \\
 \hline \hline
 Calculated From Target Density & & & \\
 \hline
$f_{\pi}$ Calculated  & 0.27  & 0.039 & \\
$f_{\pi}$ Fixed (=0.2) & 0.091 & 0.013 & \\
 \hline \hline
 Inferred From Energy Conservation & & & \\
 \hline
$E_{\gamma}^{\rm{max}}=100\,$GeV   & 0.21 & 0.065 & \\
$E_{\gamma}^{\rm{max}}=1\,  $TeV   & 2.1 & 0.65 & \\
$E_{\gamma}^{\rm{max}}=10\, $TeV   & 21. & 6.5 & \\
 \hline \hline
 \end{tabular}
 \caption{The event rate (per square kilometer) from GRB941017 for two  
calculational methods. 
Those rates found by the target density calculation 
(as described in Section 2) are given for the calculated value of
$f_\pi$ the fraction of proton energy converted to pions, 
and for a fixed value of $f_\pi=0.2$. 
This calculation assumes that roughly equal amounts of 
energy go into accelerating protons and electrons. The rates 
inferred from energy conservation are shown for three different
choices of the maximum energy to which the high-energy gamma-ray
component of the spectrum potentially extends. 
An $E^{-2}$ proton injection spectrum is used. Both muon and shower
energy thresholds are set to 100 GeV.}
 \end{table}

\end{document}